\documentclass[reprint,
groupedaddress,
unsortedaddress,
nofootinbib,
amsmath,amssymb, aps,
prl,
floatfix]{revtex4-2}

\usepackage{graphicx}
\usepackage{xcolor}
\usepackage{hyperref}
\usepackage{siunitx}
\begin{document}

\title{Quasiparticle Effective Mass of the Three-Dimensional Fermi Liquid by Quantum Monte Carlo} 

\author{Sam Azadi}
\email{s.azadi@imperial.ac.uk}
\affiliation{Department of Physics and the Thomas Young Centre for Theory and Simulation of Materials, South Kensington Campus, Imperial College London, London SW7 2AZ, United Kingdom.}
\author{N.\ D.\ Drummond}
\affiliation{Department of Physics, Lancaster University, Lancaster LA1 4YB, United Kingdom}
\author{W.\ M.\ C.\ Foulkes}
\affiliation{Department of Physics and the Thomas Young Centre for Theory and Simulation of Materials, South Kensington Campus, Imperial College London, London SW7 2AZ, United Kingdom.}
 
\date{\today}

\begin{abstract}
According to Landau's Fermi liquid theory, the main properties of the quasiparticle excitations of an electron gas are embodied in the effective mass $m^*$, which determines the energy of a single quasiparticle, and the Landau interaction function, which indicates how the energy of a quasiparticle is modified by the presence of other quasiparticles. This simple paradigm underlies most of our current understanding of the physical and chemical behavior of metallic systems. The quasiparticle effective mass of the three-dimensional homogeneous electron gas has been the subject of theoretical controversy and there is a lack of experimental data. In this work, we deploy diffusion Monte Carlo (DMC) methods to calculate $m^*$ as a function of density for paramagnetic and ferromagnetic three-dimensional homogeneous electron gases. The DMC results indicate that $m^*$ decreases when the density is reduced, especially in the ferromagnetic case. The DMC quasiparticle energy bands exclude the possibility of a reduction in the occupied bandwidth relative to that of the free-electron model at density parameter $r_\text{s}=4$, which corresponds to Na metal.
\end{abstract}

\maketitle
The fermionic many-body problem has been a major subject in physics since the early days of quantum mechanics. Its applications range from the microscale in nuclei, atoms, and molecules to the macroscale in condensed matter physics and even to the astrophysical scale, as in the physics of neutron stars \cite{Carlson,Wyk,Chin,Mazurenko}. During the past few decades the field of quantum electron liquids \cite{Landau, Pines, Giuliani, Wigner} has seen many important developments. Advances in semiconductor technology have enabled the realization of ultra-pure two-dimensional (2D) electron liquids, the densities of which can be tuned by electrical techniques, allowing a systematic exploration of both strongly and weakly correlated regimes \cite{Arp}. Experimental methods have been developed to probe with high precision the thermodynamics of locally homogeneous ultracold Bose and Fermi gases, allowing stringent tests of many-body theories \cite{Nasci10,Navon,Nasci11}. On the theoretical front, advanced computational methods such as quantum Monte Carlo (QMC) simulations have complemented approaches based on many-body perturbation theory, and have helped to place our understanding of electron liquids on a quantitative footing \cite{Ceperley80,Ceperley77,Ceperley95}.

The simplest possible electron liquid is the homogeneous electron gas (HEG), in which interacting electrons move in a uniform, inert, neutralizing background. The HEG is characterized by a very simple Hamiltonian,\footnote{We use Hartree atomic units (a.u.), in which $\hbar=4\pi\epsilon_0=m_{\rm e}=|e|=1$. The Coulomb sum in the definition of the Hamiltonian is shorthand for a convergent sum of periodic Ewald interactions.} $\hat{H}=\sum_i -(1/2)\nabla_i^2+\sum_{i>j} r_{ij}^{-1}$, which nevertheless leads to rich and challenging physical behavior \cite{Ceperley86,Foulkes,Becca,YKwon,Holzman,Spink,Booth09,Ruggeri,Shepherd,Prokof,Tanatar,Moroni,Neil_2013}. The density parameter $r_\text{s}$ plays a crucial role in the physics of the electron liquid and is given by $r_\text{s}=(\frac{3}{4n\pi})^{1/3} a_\text{B}^{-1}$, where $n$ is the electron number density and $a_\text{B}=1$ a.u.\ is the Bohr radius. 

According to Landau's theory \cite{Landau}, the main properties of a quasiparticle excitation in a Fermi liquid are embodied in its effective mass $m^*$, which determines the energy of a single quasiparticle, and its Landau interaction function $f_{k,k',\sigma}$, which tells us how the energy of a quasiparticle is modified by the presence of other quasiparticles. The quasiparticle effective mass differs in general from the bare particle mass $m_\text{e}=1$ a.u.\ and  is related to the interaction function. The quasiparticle picture of excitations applies not only to electrons in metals and doped semiconductors (where the renormalized mass $m^*$ remains close to the free electron mass $m_\text{e}$), but also to $^3$He atoms in the liquid phase, where interactions renormalize the mass by a factor of around three, and to highly correlated heavy fermion systems, where $m^*/m_\text{e}$ can run into the hundreds \cite{Stewart}. 

In this work we have used the continuum variational Monte Carlo (VMC) \cite{Umrigar,Ceperley77} and diffusion Monte Carlo (DMC) \cite{Ceperley80}  methods in real space to obtain accurate values of the 3D-HEG quasiparticle effective mass $m^*$  in the thermodynamic limit of infinite system size at different densities and for different spin polarizations. This provides us with the most important contribution to the Landau energy functional \cite{Landau,Giuliani}.

In the VMC method, parameters in a trial wave function (WF) are optimized according to the variational principle, with energy expectation values calculated by Monte Carlo integration in the $3N$-dimensional space of position vectors of the $N$ electrons.  In the DMC method, the imaginary-time Schr\"{o}dinger equation is used to evolve a statistical ensemble of electronic configurations towards the ground state. Fermionic antisymmetry is maintained by the fixed-node approximation, in which the nodal surface of the WF is constrained to equal that of an approximate WF optimized within VMC\@. The fixed-node DMC energy provides a variational upper bound on the ground state energy, with an error that is second order in the error in the nodal surface.\footnote{In fact, the DMC energy is an upper bound on the energy of the lowest-energy state that transforms as the same 1D irreducible representation of the symmetry group of the Hamiltonian as the trial WF; hence our excited state DMC energies are upper bounds on the corresponding excited states.} The energy band  ${\cal E}({\bf k})$ is calculated by evaluating the difference in the total DMC energy when an electron is added to or removed from a state with momentum ${\bf k}$. Electronic excitations close to the Fermi surface correspond to quasiparticle excitations. The electronic and quasiparticle bands therefore agree near the Fermi surface and have the same derivative at $k_\text{F}$. The effective mass of the 3D HEG can be written as $m^* = k_\text{F} / (d{\cal E}/dk)_{k_\text{F}}$, so it is straightforward to compute the effective mass once the energy band has been determined. We have studied both paramagnetic and ferromagnetic HEGs in order to investigate the difference between their behavior as a function of density. The effective mass depends on the spin polarization, and the differences in the quasiparticle effective masses of ferromagnetic and paramagnetic 3D HEGs produce different transport properties \cite{Weber}. 

Our trial WFs consist of Slater determinants of plane-wave orbitals multiplied by a Jastrow correlation factor. The Jastrow factor contains polynomial and plane-wave expansions in electron-electron separation. The orbitals in the Slater WF are evaluated at quasiparticle coordinates related to the actual electron positions by backflow (BF) \cite{Pablo} functions consisting of polynomial and plane-wave expansions in electron-electron separation. Full details are given in the Supplemental Material \cite{Suppl}. The WFs were optimized by variance minimization \cite{Neil05} followed by energy minimization \cite{Toulouse}. 

The single-particle energy for an occupied state at wavevector ${\bf k}$ is defined as ${\cal E}({\bf k}) = E_0 - E^{-}({\bf k})$, while the single-particle energy for an unoccupied state is ${\cal E}({\bf k}) =  E^{+}({\bf k}) - E_0$, where $E_0$ is the ground-state total energy, $E^{+}({\bf k})$ is the total energy of the system with an extra electron placed in orbital $\exp(i{\bf k}\cdot{\bf r})$, and $E^{-}({\bf k})$ is the total energy with an electron removed from orbital $\exp(i{\bf k}\cdot{\bf r})$. In a finite simulation cell subject to periodic boundary conditions, the available states ${\bf k}$ fall on the grid of reciprocal lattice points offset by the simulation cell Bloch vector ${\bf k}_\text{s}$ \cite{Neil08, Rajagopal94, Rajagopal95, Azadi19}. The simulation cell volume was left unchanged when electrons were added or removed. For noninteracting electron systems this gives the energy band $(k^2/2)$ without finite-size (FS) error.  Using a fixed volume is thermodynamically correct, as the energy band is defined via a difference of Helmholtz free energies. Systematic FS effects arise due to image interactions and because Friedel oscillations \cite{Dobson} are forced to be commensurate with the simulation cell. We evaluated the FS biases by performing simulations for different cell sizes with up to 274 electrons. After determining the energy band at a series of $k$ values, we performed a least-squares fit of a Pad\'{e} function ${\cal E}(k)=(A_0+A_1k+A_2k^2+A_3^2k^3)/(1+2A_3^2k)$ to the band values \cite{Suppl}. 

The DMC energy band is defined via differences in total-energy eigenvalues; these differences coincide with the quasiparticle band near the Fermi surface and hence give a correct description of the effective mass. The number of electrons $N$ in each of our ground-state calculations was chosen to be a ``magic number'' corresponding to a closed-shell configuration when ${\bf k}_\text{s}={\bf 0}$. It is therefore appropriate to use real, single-determinant, WFs for the ground-state calculations. In the $(N+1)$- and $(N-1)$-electron open-shell excited-state calculations, we used the Jastrow factor and BF function that were optimized for the $N$-electron ground state in a complex trial WF\@.

\begin{figure*}[t]
    \centering
    \begin{tabular}{c c c}
    \includegraphics[scale=0.26]{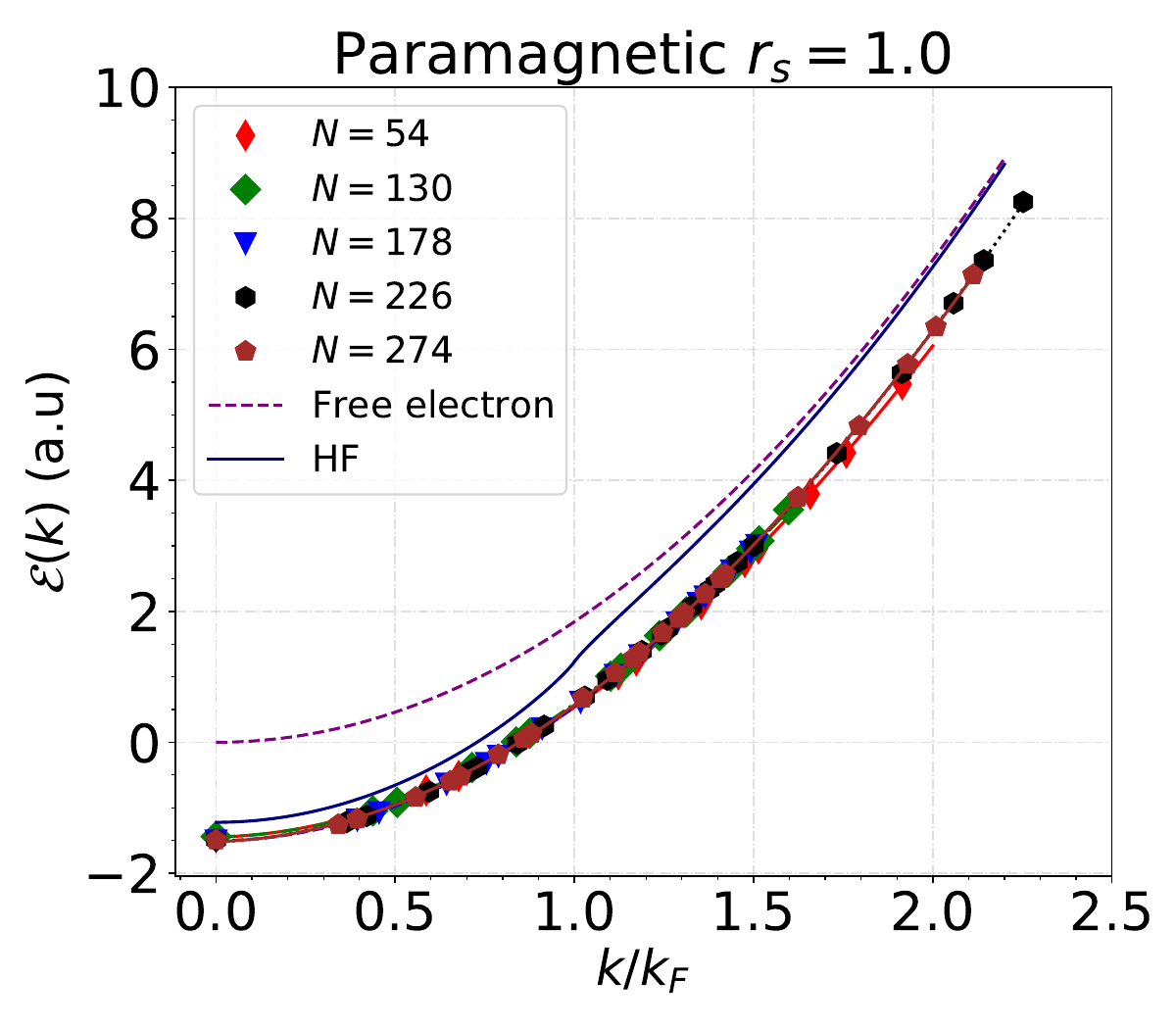}& 
    \includegraphics[scale=0.26]{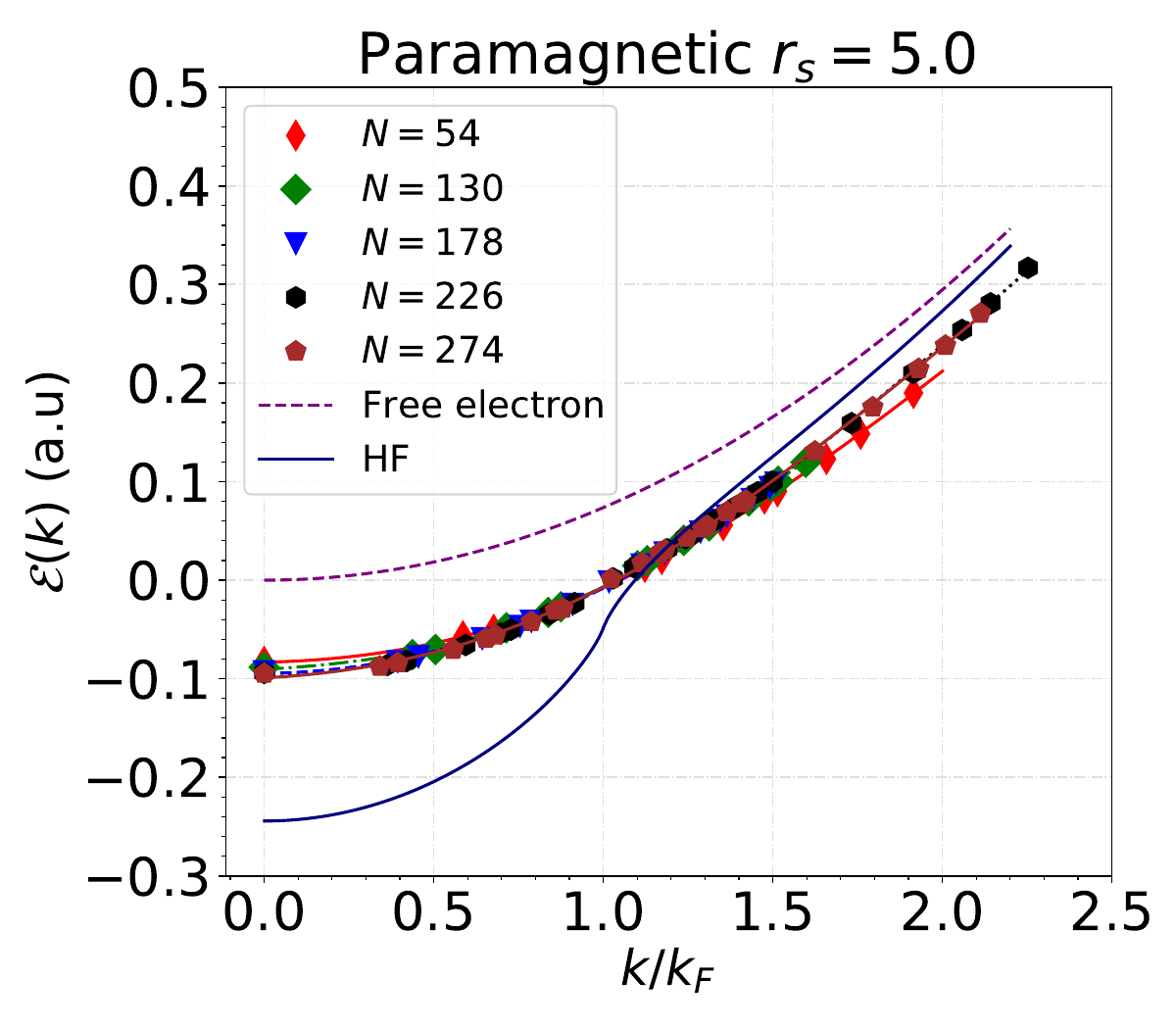}& 
    \includegraphics[scale=0.26]{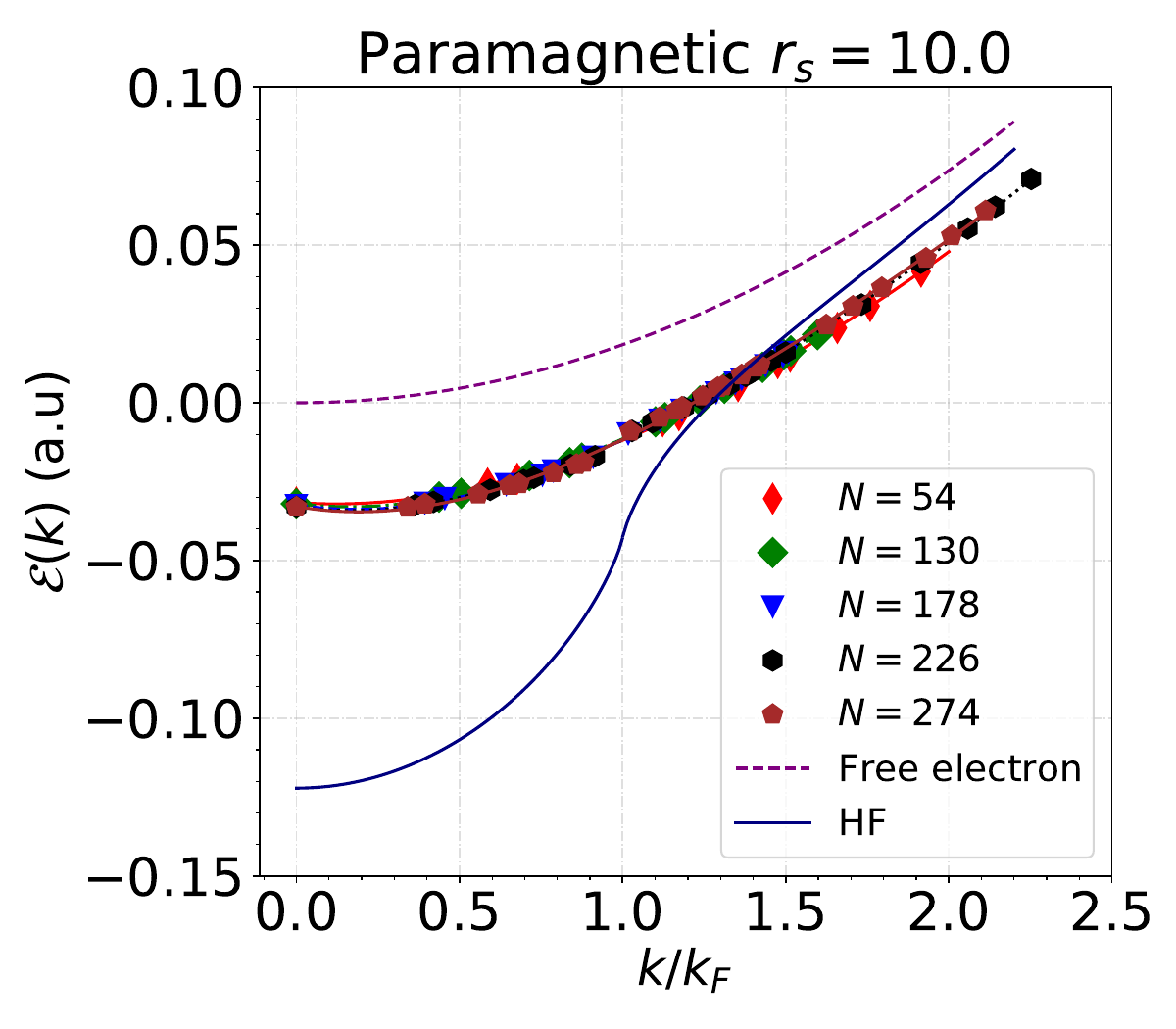}\\
    \includegraphics[scale=0.26]{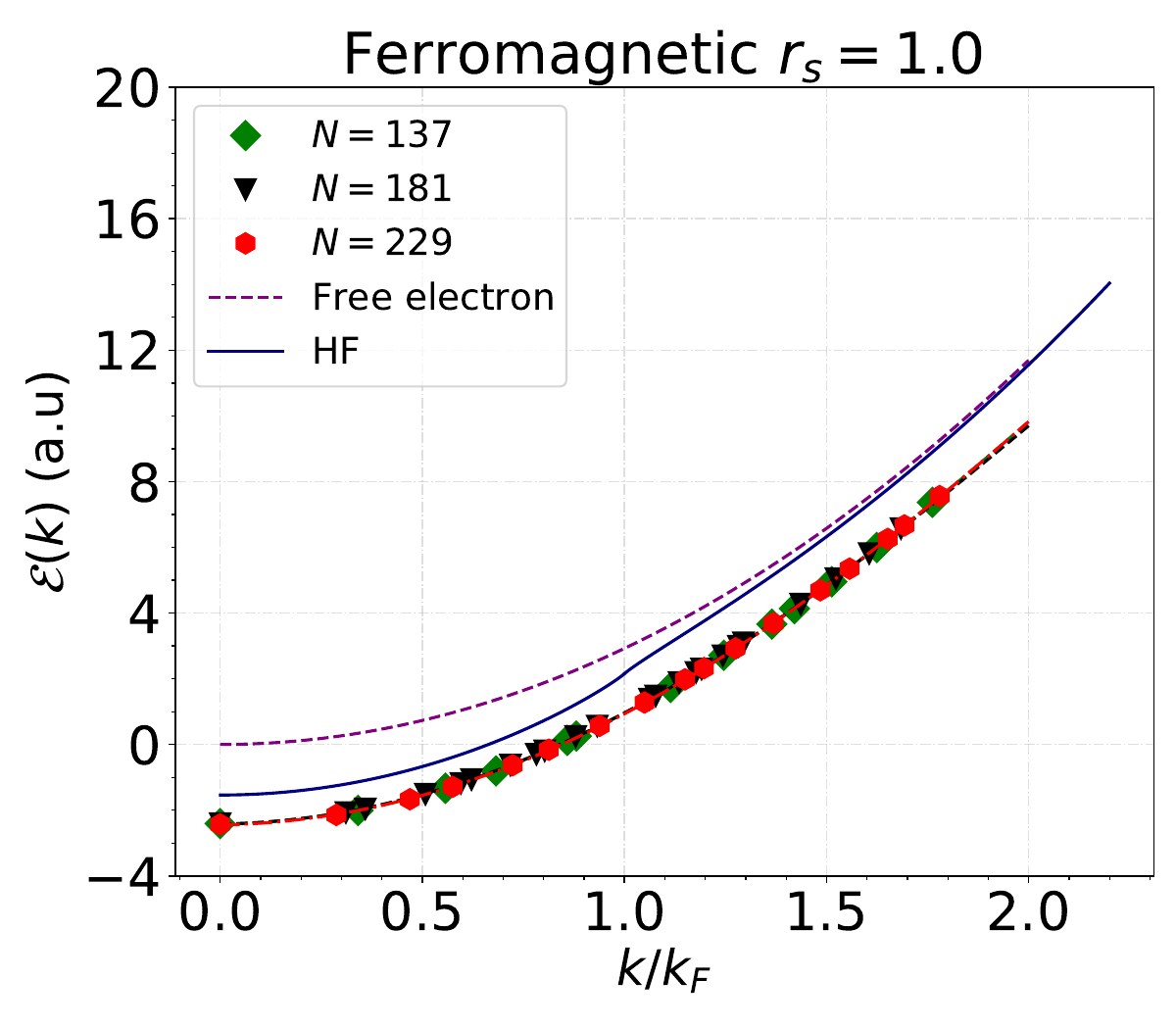}&
    \includegraphics[scale=0.26]{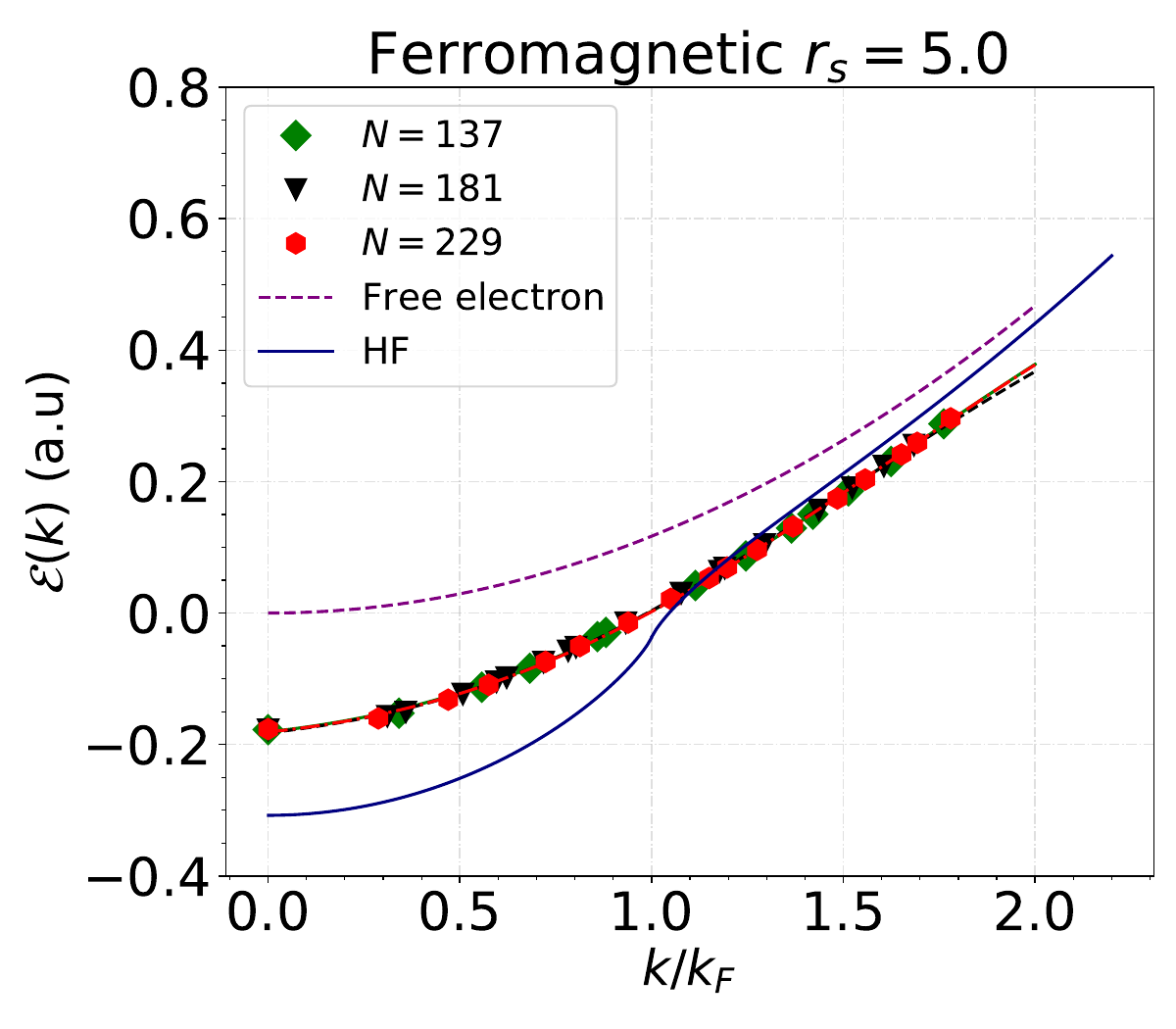}&
    \includegraphics[scale=0.26]{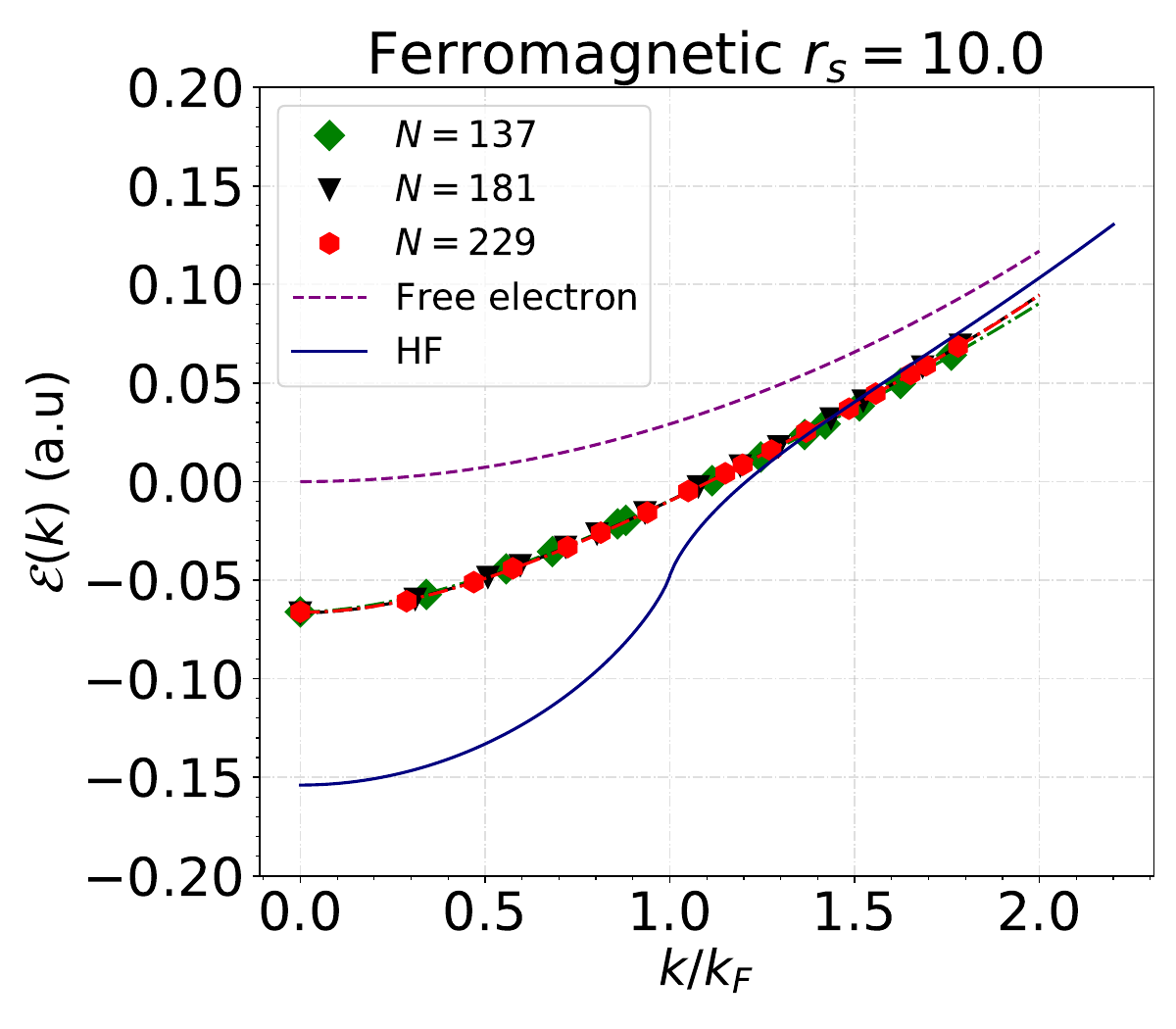}\\
    \end{tabular}
    \caption{\label{bands} (First row) DMC energy bands for paramagnetic 3D-HEGs at $r_\text{s}=1$, 5, and 10 obtained using system sizes of $N=54$, 130, 178, 226, and 274 electrons. Pad\'{e} functions are fitted to the DMC data. (Second row) DMC energy bands for ferromagnetic 3D HEGs at $r_\text{s}=1$, 5, and 10, obtained using system sizes of $N=137$, 181, and 229 electrons. Pad\'{e} functions are fitted to the DMC data. DMC energy bands for paramagnetic and ferromagnetic 3D-HEGs at $r_\text{s}=2$, 3, and 4 are plotted in the Supplemental Material \cite{Suppl}. }
\end{figure*}

We studied the 3D-HEG at six different densities, $r_\text{s}=1$, 2, 3, 4, 5, and 10. For each density, we performed QMC calculations for simulation cells containing $N = 54$, 130, 178, 226, and 274 electrons for paramagnetic HEGs and $N = 137$, 181, and 229 electrons for ferromagnetic HEGs. To obtain the full energy band, we calculated the quasiparticle energies at up to twenty momenta within the range $0 \leq k < 2k_\text{F}$. In the thermodynamic infinite-system-size limit, the exact energy band is smooth and well-behaved at the Fermi surface. However, the HF band is pathological (see Fig.\ \ref{bands}) and has a logarithmic divergence at the Fermi surface in the thermodynamic limit \cite{Giuliani}. In finite systems, the HF band oscillates wildly at wavevectors near the Fermi surface, and although DMC retrieves a large fraction of the correlation energy, it does not entirely eliminate the pathological behavior inherited from HF theory \cite{Neil_2013}. Hence, it is necessary to consider excitations away from the Fermi surface to obtain the gradient of the energy band at $k_\text{F}$. 

The HF nodal surface of an electron system is in general both quantitatively and qualitatively (topologically) incorrect. Backflow functions do not change the nodal topology, and the comparison of DMC results with and without backflow (see the Supplemental Material \cite{Suppl}) suggests that quantitative nodal errors in our DMC effective masses are small.  The issue of qualitative errors in the nodal surface is closely related to the question of whether or not Landau's Fermi liquid theory is valid.  If Fermi liquid theory is valid then the topology of at least those sections of the nodal surface that are relevant to low-lying excitations must be the same as the topology of the HF wave function (a Slater determinant of plane waves), because of the assumed adiabatic connection between interacting and noninteracting electron gases. We therefore believe fixed-node errors in our effective masses to be small.

The \textsc{casino} package was used for all our QMC calculations; this software has previously been used in numerous studies of HEGs \cite{casino}. For the calculation of the effective mass of the 3D-HEG the WF includes the polynomial two-body Jastrow term $u$, the polynomial two-body BF term $\eta$, the plane-wave two-body Jastrow term $p$, and the plane-wave two-body BF term $\pi$. The VMC and DMC energies and VMC variances of 3D paramagnetic HEGs obtained using different WFs are presented and discussed in the Supplemental Material \cite{Suppl}.

The DMC energy bands of paramagnetic and ferromagnetic 3D-HEGs with density parameters of $r_\text{s}=1$, 5, and 10 are illustrated in Fig.\ \ref{bands}. The DMC energy bands for densities of $r_\text{s}=2$, 3, and 4 are reported in the Supplemental Material \cite{Suppl}. 
The free-electron and HF bands are also shown in Fig.\ \ref{bands}. The noninteracting free-electron band is much more accurate than the HF band, particularly in the low density regime. The pathological behavior of the HF band is driven by the long range of the exchange hole producing an incomplete screening of the Coulomb interaction \cite{Giuliani}; the exchange-correlation hole, by contrast, falls off much more rapidly \cite{gorigiorgi_perdew}. The occupied bandwidths of paramagnetic and ferromagnetic 3D-HEGs at different density parameters $r_\text{s}$ obtained using the DMC, free-electron, and HF methods are illustrated in Fig.\ \ref{BW}. The DMC bandwidth values are also listed in the Supplemental Material \cite{Suppl}. The DMC bandwidths of paramagnetic and ferromagnetic 3D-HEGs are larger than the corresponding free-electron bandwidths but smaller than the HF bandwidths at each density studied. The paramagnetic DMC bandwidth at $r_\text{s}=4$ (the appropriate density parameter for Na metal) agrees with the bandwidth of Na obtained using QMC and many-body $GW$ calculations \cite{Maezono, Yasuhara, Takada}.
\begin{figure}[!htbp]
    \centering
    \includegraphics[scale=0.32, angle=0]{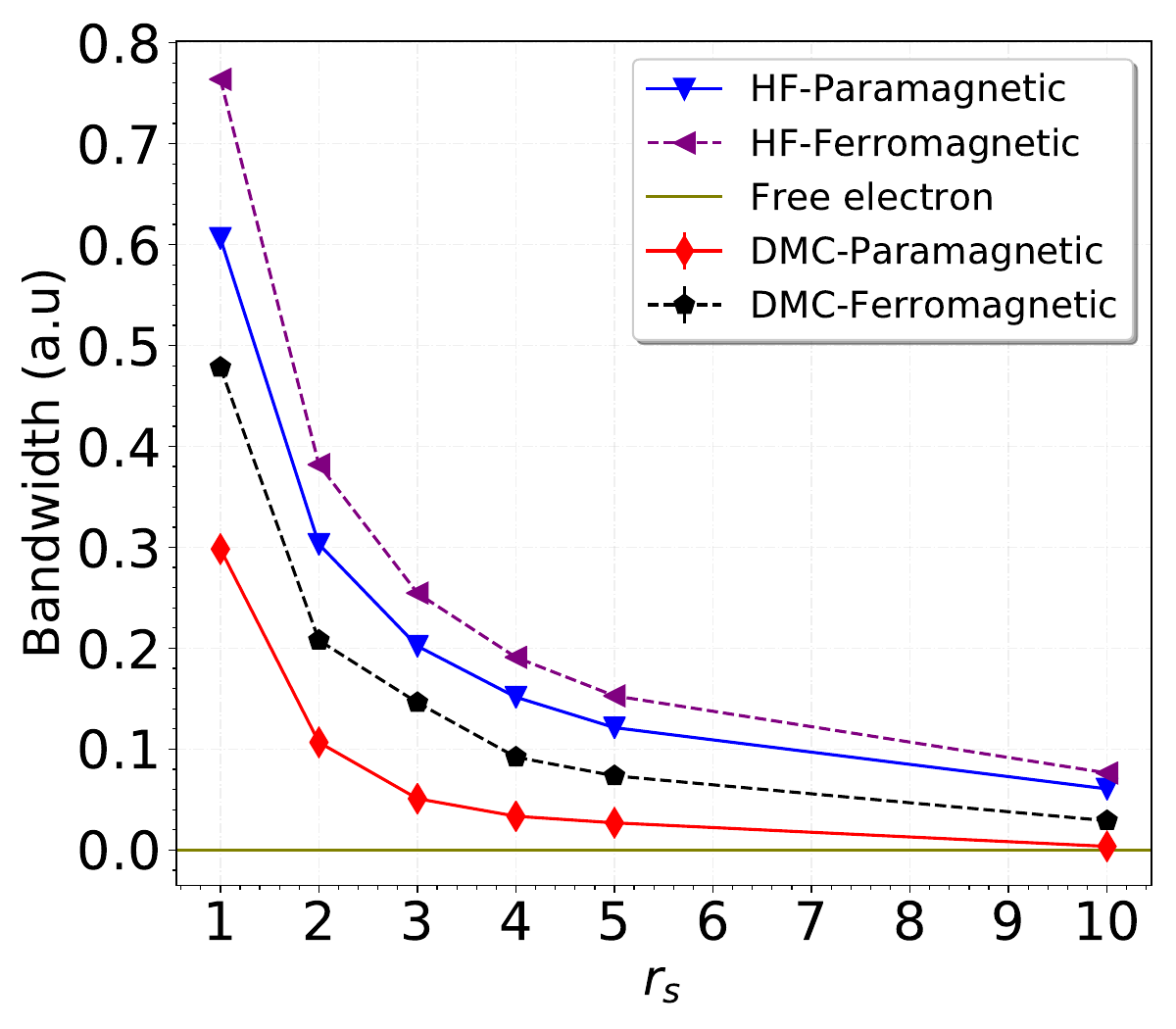}\\
    \caption{\label{BW} Occupied bandwidths of paramagnetic and ferromagnetic 3D-HEGs as a function of the density parameter $r_\text{s}$ calculated using the HF and DMC methods. All bandwidths are given relative to the corresponding free-electron value.}
    \label{m_rS1}
\end{figure}

The quasiparticle effective masses of paramagnetic and ferromagnetic 3D-HEGs, obtained from the Pad\'{e} fits of the energy bands using the equation $m^* = k_\text{F} / (d{\cal E}/dk)_{k_\text{F}}$, are plotted against system size in Fig.~\ref{FS}. Similar results obtained using quartic fitting functions are reported in the Supplemental Material \cite{Suppl}. The effect of using different fitting functions on $m^*$ is negligible, especially for ferromagnetic systems. We also used a Slater-Jastrow (SJ) WF to calculate $m^*$ for the paramagnetic case with system size $N=178$ and density parameters of $r_\text{s}=1$ and 10 and we found that both SJ and Slater-Jastrow-backflow (SJB) WFs yield similar effective masses \cite{Suppl}. In the case of $r_\text{s}=1$ the values of $m^*$ obtained by SJ and SJB WFs are 0.915(1) and 0.921(1), respectively, and for density parameter of $r_\text{s}=10$ the values of $m^*$ calculated by SJ and SJB WFs are 0.75(1) and 0.78(1), respectively \cite{Suppl}. Because of a systematic trend in the effective mass as a function of system size, we extrapolated $m^*$ to the thermodynamic limit for each density and spin polarization. For paramagnetic HEGs, the extrapolation reduces $m^*$ more when $r_\text{s}=5$ and $10$ than at higher densities. 
\begin{figure}[!htbp]
    \centering
    \begin{tabular}{c}
    \includegraphics[scale=0.32, angle=0]{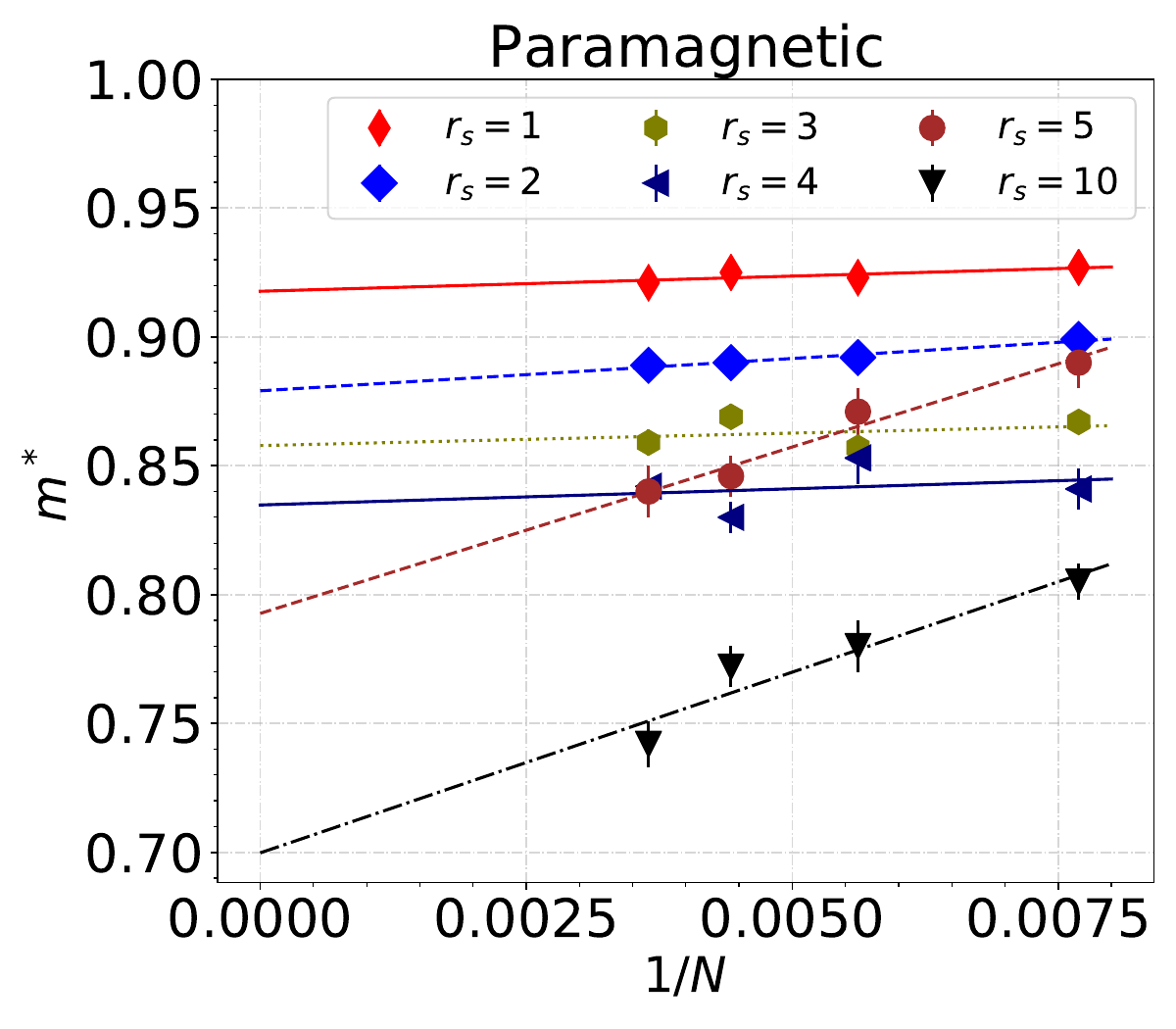}\\
    \includegraphics[scale=0.32, angle=0]{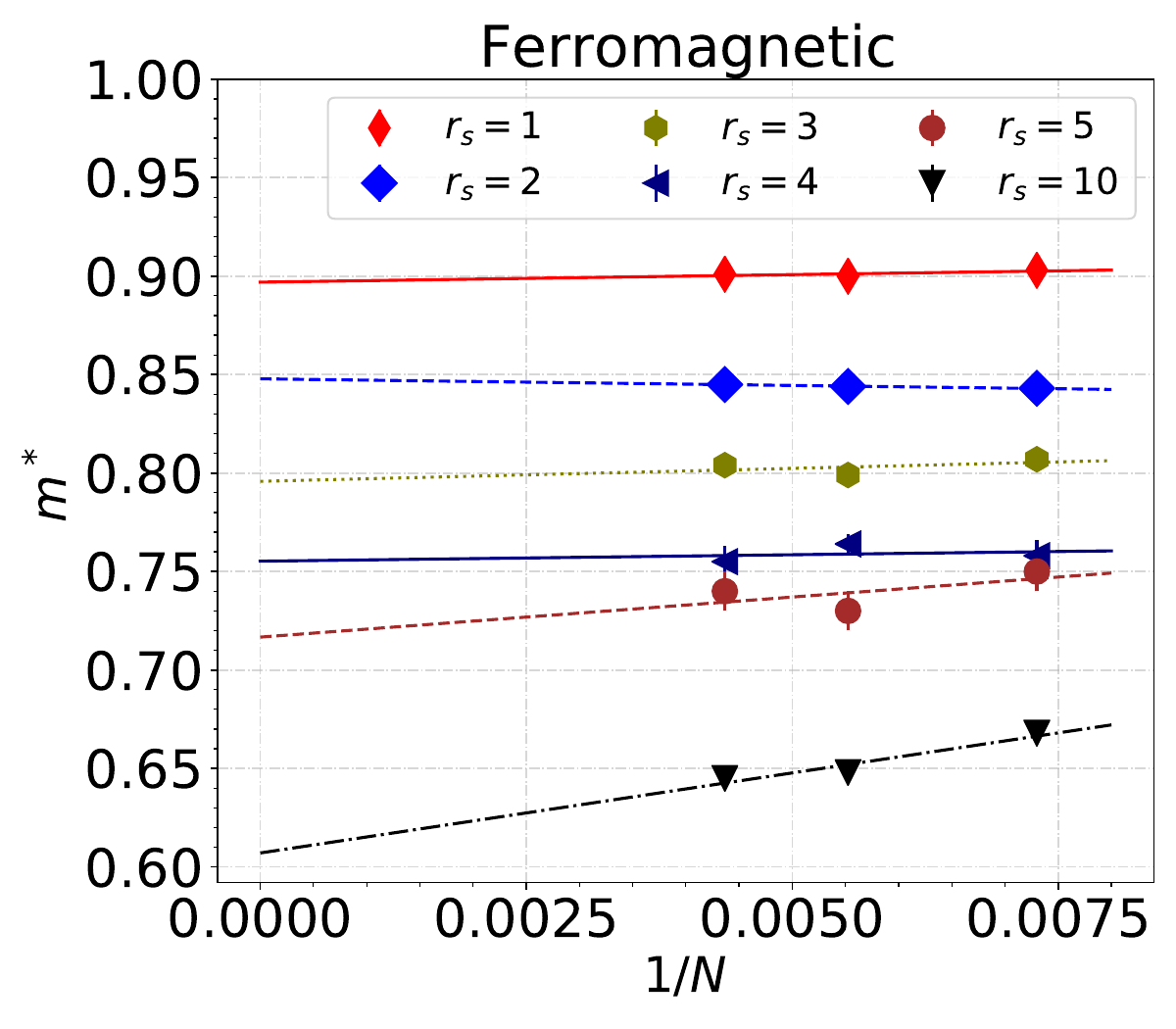}\\
    \end{tabular}
    \caption{\label{FS} Quasiparticle effective masses $m^*$ of paramagnetic and ferromagnetic 3D-HEGs as functions of $1/N$, where $N$ is the system size. The Monte Carlo statistical error bars are included, but they are smaller than the symbols. It should be noted that these error bars only quantify the noise in our data due to the Monte Carlo process.  Unquantified noise includes: (i) the effects of stochastically optimized backflow functions on the fixed-node DMC energy; and (ii) quasirandom finite-size effects, such as Friedel oscillations being forced to be commensurate with the simulation cell.}
    \label{m_rS2}
\end{figure}
\begin{figure}[!htbp]
    \centering
    \includegraphics[scale=0.35]{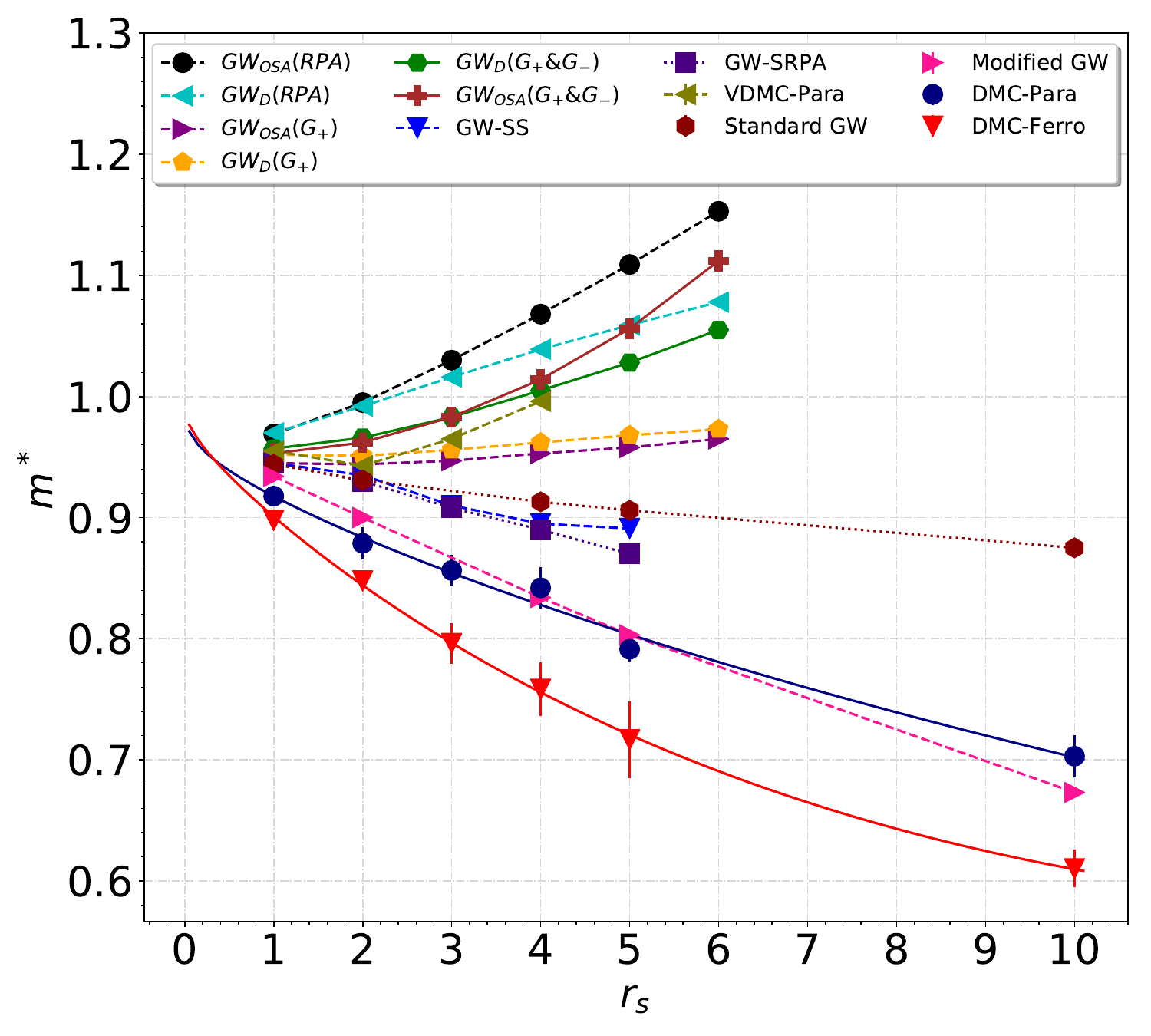}\\
    \caption{\label{eff_mass} Quasiparticle effective masses $m^*$ of paramagnetic (Para) and ferromagnetic (Ferro) 3D-HEGs at the infinite-system-size limit as functions of density parameter $r_\text{s}$. Pad\'{e} functions were fitted to the DMC quasiparticle energy bands to determine the effective mass. 
    The many-body $GW_x$ and variational diagrammatic Monte Carlo (VDMC) results are from Refs.\ \cite{Simion} and \cite{Haule}, respectively. The $GW$-SS and $GW$-SRPA results are from Refs.\ \cite{Krakovsky} and \cite{Rietschel}, respectively. The ``standard $GW$'' data and ``modified $GW$'' data are taken from Ref.\ \cite{Houcke}. All the $GW$ results are for paramagnetic 3D-HEGs. }
\end{figure}

The DMC effective masses of paramagnetic and ferromagnetic 3D-HEGs in the thermodynamic limit are plotted against density in Fig.\ \ref{eff_mass}. The difference between the paramagnetic and ferromagnetic effective masses increases at low density, as is also the case in 2D \cite{Neil_2013}. Our results indicate that the effective masses of both paramagnetic and ferromagnetic 3D-HEGs decrease when the density is reduced. The ferromagnetic $m^*$ drops faster and is smaller than the paramagnetic $m^*$ at all densities studied here. By contrast, in the 2D-HEG, $m^*$ decreases when the density is lowered in the ferromagnetic state but increases in the paramagnetic state \cite{Neil_2013}. 

The contradictory results of many-body perturbation theory calculations of the behavior of $m^*$ as a function of density have given rise to a controversy. On one side, a number of many-body perturbation theory calculations using different Green's function flavors and a variety of approximations indicate that the effective mass of the 3D-HEG increases at low density \cite{Simion}. On the other side, the use of the $GW$ approximation with the Sj\"{o}lander-Stott (SS) theory of the two-component plasma \cite{Krakovsky}, and $GW$ calculations with a random-phase-approximation-screened free-electron model (SRPA) \cite{Rietschel}, suggest that the effective mass decreases at low density.  The fully self-consistent $GW$ results for the paramagnetic 3D-HEG are significantly improved by enforcing the particle-number conservation law in the polarization function \cite{Houcke}. The effective masses predicted by this ``modified $GW$'' technique agree well with our DMC results for the whole range of densities studied. 
The $GW$ approximation is expected to be accurate at high density ($r_\text{s} \leq 1$), which is consistent with the behavior shown in Fig.\ \ref{eff_mass}, where the differences between the various $GW$ results reduce as the density increases.  Indeed, the difference between the DMC and $GW$ effective masses is quite small at $r_\text{s}=1$. Recently, the single-particle excitation spectra and quasiparticle effective masses of 3D-HEGs have been calculated using variational diagrammatic Monte Carlo (VDMC) \cite{Haule}, in which high-order Feynman diagrams are sampled using Monte Carlo methods \cite{Chen}. The behavior of the VDMC effective mass as a function of density is close to some of the $GW$ results, as can be observed from Fig.\ \ref{eff_mass}.  To the best of our knowledge, there are no reliable experimental results for the effective mass of the 3D-HEG\@. However, the bandwidth of Na metal, which has a band effective mass (incorporating crystal lattice effects) of 1.23, has been measured \cite{Jensen, Lyo} and can be compared with that of the 3D-HEG at density parameter $r_\text{s}=4$. Neither our DMC results nor the existing VDMC and $GW$ results explain the experimentally estimated 18--25\% bandwidth narrowing relative to self-consistent band theoretical calculations \cite{Jensen, Lyo}. 

In summary, we have calculated the single-particle energy bands and quasiparticle effective masses of paramagnetic and ferromagnetic 3D-HEGs using the DMC method. Two fitting functions, of Pad\'{e} and quartic form, have been used to obtain the gradient of the energy band at the Fermi wavevector and hence the effective mass at each finite system size studied. We found that the effective masses of paramagnetic and ferromagnetic systems of any given finite size are almost independent of the choice of trial WF and the fitting function used. The DMC bandwidths of paramagnetic and ferromagnetic 3D-HEGs are larger than that of the free-electron model but smaller than the HF bandwidth at all densities considered. The DMC bandwidth for a 3D-HEG with density parameter $r_\text{s}=4$ agrees with previous QMC results for the bandwidth of Na. A sufficiently high precision is achieved in our simulations that the systematic finite-size errors in the effective masses can be eliminated by extrapolation to the thermodynamic limit. Our DMC results predict that the effective mass of the 3D-HEG decreases as the density decreases from $r_1 = 1$ to $r_\text{s} = 10$. This reduction is more pronounced in the ferromagnetic system than the paramagnetic system. The good agreement between DMC results for Na and the 3D-HEG indicates that the 3D-HEG provides a good model. Therefore, the considerable decrease of Na bandwidth suggested by experiment could be due to surface or lattice effects \cite{Shung}. 

We acknowledge support from the Thomas Young Centre under Grant number TYC-101. S.\ Azadi acknowledges PRACE for awarding us access to the High-Performance Computing Center Stuttgart, Germany, through the project 2020235573.


\end{document}